\documentclass[11pt]{article}
\usepackage[utf8]{inputenc}
\usepackage[english]{babel}
\usepackage{siunitx}
\usepackage[margin=1in]{geometry}
\usepackage{amsmath, amssymb, amsfonts}
\usepackage{mathtools}
\usepackage{amsthm}
\usepackage[export]{adjustbox}
\usepackage{graphicx}
\usepackage{subcaption}
\usepackage{caption}
\usepackage{booktabs}
\usepackage{float}
\usepackage{algorithm}
\usepackage{algpseudocode}
\usepackage[numbers]{natbib}
\usepackage{xurl}
\usepackage{hyperref}
\usepackage{doi}
\usepackage[export]{adjustbox}
\title{Assessing the Information Content of Individual Spikes in Population-Level Models of Neural Spiking Activity}

\author{
Azar Ghahari\thanks{Department of Mathematics and Statistics, Boston University, Boston, MA 02215, USA.} \and
Uri T. Eden\footnotemark[1]
}

\date{}

\begin{document}

\maketitle

\begin{abstract}
In the last decade, there have been major advances in clusterless decoding algorithms for neural data analysis. These algorithms use the theory of marked point processes to describe the joint activity of many neurons simultaneously, without the need for spike sorting. In this study, we examine information-theoretic metrics to analyze the information extracted from each observed spike under such clusterless models. In an analysis of spatial coding in the rat hippocampus, we compared the entropy reduction between spike-sorted and clusterless models for both individual spikes observed in isolation and when the prior information from all previously observed spikes is accounted for. Our analysis demonstrates that low-amplitude spikes, which are difficult to cluster and often left out of spike sorting, provide reduced information compared to sortable, high-amplitude spikes when considered in isolation, but the two provide similar levels of information when considering all the prior information available from past spiking. These findings demonstrate the value of combining information measures with state-space modeling and yield new insights into the underlying mechanisms of neural computation.
\end{abstract}


\section{Introduction}

Neural decoding analyses serve a multitude of purposes. When the signals being decoded are observable, decoding can help validate the quality of neural encoding models and reveal how neural systems integrate information from upstream areas \cite{Brown1998}. When signals are not observable, decoding can provide insight into the internal cognitive processes underlying complex brain states \cite{Chen2014, Zoltowski2020ML, Denovellis2021}. One feature that distinguishes decoding from encoding analyses is that encoding models have historically focused on the receptive field properties of individual neurons, while decoding, by its nature, is typically focused on integrating information across multiple neurons. Advances in decoding approaches often aim to maximize the use of all available information from the recorded neural population on the represented signals \cite{Kloosterman2013}. Even when improved decoding methods are developed, it often remains a mystery how those methods better extract information from the observed data. 

Many neural decoders that work on spiking populations rely on a preliminary spike sorting step \cite{Quiroga2004} to identify the set of neurons responsible for each recorded waveform \cite{VargasIrwin2007}. However, spike sorting in the presence of noisy recordings is challenging and while sorting algorithms have become more advanced in recent years \cite{Lewicki1998}, different approaches often show large variability in the estimated number of neural sources and the clustering of each waveform {\cite{Lewicki1998,Wood2004,Einevoll2012,Fee1996}. Imperfect spike sorting is known to lead to bias in estimating receptive fields, and bias and increased variability in decoding \cite{Ventura2009}. Additionally, most spike sorting methods discard spikes that cannot be clustered with high confidence. This includes many low amplitude spikes that do not form clean separable clusters.  Recent advances in extracellular recording techniques enable the observation of activity from thousands of neurons simultaneously, facilitating the study of large neuronal networks but also increasing the dimensionality and complexity of the recorded signals \cite{Paulk2022,Lefebvre2016}. The computational complexity of spike sorting increases significantly as neuroscience experiments incorporate larger and more densely populated electrode arrays \cite{Todorova2014}.

Clusterless decoding models avoid the need for spike sorting by using all recorded spikes directly without assigning them to specific neurons \cite{Kloosterman2013}. This method, which relies on the timing and waveform features of spikes, has proven particularly effective in analyzing large datasets where spike sorting is impractical. A successful approach has used the theory of marked point processes to model all observed spikes from a population and their waveforms as a function of the encoded signals \cite{Kloosterman2013,Deng2015}. These clusterless point process models have been broadly applied to the problem of estimating movement trajectories and nonlocal spatial representations of the hippocampi of rats performing spatial navigation tasks \cite{Denovellis2021,Kloosterman2013,Deng2015,Deng2016,Tao2018,Kay2020,Gillespie2021} and have been shown to lead to better location estimates and more accurate confidence bounds on these estimates compared to decoding from sorted spikes \cite{Deng2015}. However, the precise factors that contribute to this improvement remain unclear.

One potential factor is the reduction of bias due to spike sorting errors, which propagates to encoding models and decoded estimates \cite{Ventura2009}. Another potential factor is the information that is lost in spike sorting when spikes cannot be clustered and are therefore omitted from decoding analyses. This often includes low amplitude spike events from distal neurons, which are large enough to trigger a spike detection threshold, but which have waveforms that are overshadowed by recording noise. However, it is unclear whether useful information can be extracted from the inclusion of such spikes in decoding analyses. More research is needed to determine whether the improvement observed from clusterless decoders arises from their handling of unsorted spike data, their ability to capture a broader spectrum of neural activity patterns, their use of advanced computational algorithms, or a combination of these factors.

Information theory \cite{shannon1948} provides a framework for quantifying the amount of information neural responses and a corresponding neural model convey about different stimuli by assessing their statistical variability \cite{Borst1999}. For neural data analysis, a useful set of metrics are entropy and differential entropy, which are measures of uncertainty that can represent the information that is still needed to describe a stimulus when accounting for the information the decoder has extracted from the neural responses. The reduction in entropy explains the decrease in uncertainty, quantifying the information gained when observing a specific neural response or spike train. A high reduction in entropy suggests that the neural model effectively captures key stimulus features or patterns \cite{Hurwitz1975, Rieke1997}.  

This motivates us to explore the information extracted from individual spikes during decoding. We implemented a Bayesian filter that computes the posterior distribution of the signal represented at each time point as a function of the observed spiking activity to that time \cite{Kloosterman2013, Eden2004}. The posterior represents our state of knowledge about the signal based on all the spikes used for decoding, including our estimate of the signal and uncertainty about that estimate. We examined the influence of individual spikes on the posterior distributions based on two measures: the reduction in entropy (RE) and the reduction in the root mean squared error (RrMSE). The reduction in entropy characterizes the degree to which each spike makes our estimate of the signal more certain, and the reduction in root mean squared error describes the accuracy of this information. 

We applied these measures to a decoding analysis of the movement of a rat performing a memory-guided spatial navigation task based on population spiking data from its hippocampus. We compared the reduction in entropy and root mean squared error between spike-sorted and clusterless models of the same spike, and between clusterless models of those spikes that were included in a spike-sorted decoding analysis and those that were excluded because they could not be confidently clustered. We also compared these measures when each spike was considered on its own, without including prior information from previous spiking, and the information provided by each spike when all previous spiking was taken into account. This allowed us to differentiate the maximal information that each spike could provide from the incremental information that it actually provided during the decoding process.

The remainder of this paper is organized as follows. In Section 2, we review methods for clusterless encoding and decoding and define the measures used to quantify the information obtained from each spike during decoding. Section 3 applies these methods to population spiking data from rat hippocampus as the animal performed a memory guided spatial navigation task, and uses them to investigate the reasons that clusterless decoding leads to more accurate and precise estimates than spike-sorted decoding. Section 4 discusses the broader implications of our findings, including their applicability to other datasets and experimental settings.

\section{Methods}

Our primary objective is to quantify the information content of individual spikes,  both in isolation and in relation to prior spiking activity. To accomplish this, we use a Bayesian state-space modeling framework \cite{Eden2004} to compute a prior distribution of the signal at each spike time.  This prior is then updated based on the likelihood of the observed spike as a function of that signal. We apply information theoretic measures to compare the prior distribution of the signal before the spike with the posterior distribution after incorporating information from the most recent spike.

\subsection{Modeling methods and information measures}

Each spike information measure is computed from a pair of distributions. Let \( X_t \) be the random signal to be decoded, with \( f_{\mathrm{prior}}(X_t) \) denoting the prior distribution of \( X_t \) before observing the spike, and \( f_{\mathrm{post}}(X_t) \) the posterior distribution after observing it.

The first measure we consider is the reduction in differential entropy between the prior and posterior:
\begin{equation}
\mathrm{RE}(X_t) = -\int_{X_t} f_{\mathrm{prior}}(X_t) \log f_{\mathrm{prior}}(X_t) \, dX_t + \int_{X_t} f_{\mathrm{post}}(X_t) \log f_{\mathrm{post}}(X_t) \, dX_t.
\end{equation}
The reduction in entropy (RE) quantifies the extent to which the spike increases the concentration of the distribution, thereby reducing uncertainty in the estimate of the signal. However, this measure does not differentiate between accurate and inaccurate estimates. To assess the accuracy of the information provided by individual spikes, we also examine the reduction in absolute error (RAEr).

\begin{equation}
\mathrm{RAEr}(X_t) = \left| \int_{X_t} f_{\mathrm{prior}}(X_t) X_t dX_t -X_{t,\mathrm{true}} \right| - \left| \int_{X_t} f_{\mathrm{post}}(X_t) X_t dX_t -X_{t,\mathrm{true}} \right|.
\end{equation}
Here, $X_{t,\mathrm{true}}$ is the true value of the signal that is decoded at time $t_k$. This reduction in error measures the degree to which the observed spike improves an estimate based on the expected value of $X_t$. Finally, we examine the reduction in the root mean squared error (RrMSE),

\begin{equation}
\mathrm{RrMSE}(X_t) = \sqrt{\int_{X_t} f_{\mathrm{prior}}(X_t) (X_t-X_{t,\mathrm{true}})^2 dX_t} - \sqrt{\int_{X_t} f_{\mathrm{post}}(X_t) (X_t-X_{t,\mathrm{true}})^2 dX_t}.
\end{equation}
This measure is large when the observed spike makes the estimate both more accurate and certain. Note that we have expressed each of these measures as integrals, under the assumption that the prior and posterior distributions are defined by continuous densities. If the signal is discrete or has been discretized, and the distributions are instead given by probability mass functions, equivalent measures result from replacing the integrals with sums over full set of discrete signal values. 

The terms prior and posterior suggest Bayesian estimation, and indeed, we use a state space model structure and iteratively apply the Bayes rule to estimate $f_{\mathrm{prior}}(x_t)$ and $f_{\mathrm{post}}(x_t)$ at each time step. For a spike that occurs at time $t$, we compute, $L(x_t)$, the likelihood of observing that spike as a function of any of the possible values of $x_t$. Then, given the prior distribution, $f_{\mathrm{prior}}(x_t)$, we compute the posterior distribution, $f_{\mathrm{post}}(x_t)$, as   
\begin{equation*}
f_{\mathrm{post}}(x_t) \propto f_{\mathrm{prior}}(x_t) L(x_t),
\end{equation*}
and normalize $f_{\mathrm{prior}}(x_t)$ so that it integrates to 1.

We compare the likelihoods of two spiking models, one that uses data that have previously been spike sorted to define separate receptive fields for each sorted cluster, and one that uses unsorted, or clusterless, data to define the joint coding properties of a population of simultaneously recorded neurons \cite{Deng2015}. 

For the spike sorted model, we define a distinct receptive field model for each neuron as a function of $x_t$. Let $\Lambda^j(x_t)$ be the rate function of an inhomogeneous Poisson process that describes the probability of observing a spike from the neuron $j$ in a small interval around time $t$. The likelihood of observing a spike from that neuron at time $t$ is then given by $L_{\mathrm{sorted}}(x_t) = \Lambda^j(x_t)$.

For the clusterless spiking model, we define a single joint mark intensity function \cite{Deng2015}, $\lambda(x_t,m)$, which describes the joint probability distribution of seeing a spike in an interval around time $t$ and of that spike having specific waveform features. Here, $m$ is called the mark process, which describes a set of features of the spike waveform. This set of features could be high dimensional (e.g. a discretized trace of the waveform) or low dimensional (e.g. the peak amplitude of the spike on each of four electrodes comprising the recording tetrode). The likelihood of observing a spike with waveform features defined by $m$ at time $t$ is then given by $L_{\mathrm{clusterless}}(x_t) = \lambda(x_t,m)$. Note that the likelihood of the sorted model $L_{\mathrm{sorted}}(x_t) = \Lambda^j(x_t)$ and the likelihood of the clusterless model $L_{\mathrm{clusterless}}(x_t) = \lambda(x_t,m)$ are not directly comparable to each other, since the clusterless model describes the additional features of the data (i.e., the spike waveform structure). However, the information these likelihoods provide about the estimation of $x_t$ using the information measures described above is comparable.

To assess the information content of each spike, we consider two scenarios by selecting different prior distributions, $f_{\mathrm{prior}}(X_t)$. In the first scenario, we select a prior that provides minimal or no information about \( X_t \) to determine solely the maximum amount of information that can be obtained from the spike. We select the maximum entropy distribution for \( X_t \). Since \( X_t \) is defined on a bounded track, this is a uniform distribution, \( f_{\mathrm{prior}}(x_t) = c \), where \( c \) is the constant ensuring that the distribution integrates to 1. We refer to the resulting reduction in information measures as the \textit{isolated information} that the spike provides about the rat's location. This framework for defining the prior follows from the work of Brenner et al. \cite{brenner2000synergy}.

 In contrast to this non-informative prior, we define a prior that incorporates knowledge about the dynamics of $X_t$ and the information provided by previous spikes using a point process filter \cite{Eden2008}. Each time a new spike is observed, the filter distribution is updated, and this information is carried forward based on the state evolution model to the moment before the spike is evaluated. This becomes the prior distribution with updates based on the likelihood of that spike. 

We define a state model to capture the dynamics of the signal over time as a Wiener-driven stochastic process, 
\begin{equation} \label{eq:SS}
dx_t = Ax_t+Bdw_t,
\end{equation}
where $w_t$ is a standard Wiener process, and $A$ and $B$ are constant matrices defining the drift and scale parameters of the process, respectively. We have previously shown \cite{Eden2008} that when such a process is observed through an ensemble of spiking neurons with receptive fields $\Lambda^j_{x_t}$, in any interval with no spikes, the filter distribution evolves according to the partial differential equation,
\begin{equation} \label{eq:FPD}
\frac{\partial \rho}{\partial t} = -A\rho-Ax_t \frac{\partial \rho}{\partial x_t} + \frac{1}{2}BB^T \frac{\partial^2\rho}{\partial x^2_t}-\rho \sum^C_{j=1}{\Lambda^j(x_t)},
\end{equation}
where $\rho$ is shorthand for $\rho(t,x_t)$, the filter solution providing the density of $x_t$ at each time and $C$ is the number of observed neurons. Similarly, when the process is observed through a marked point process with joint mark intensity $\lambda(x_t,m)$, in any interval with no spikes, the filter distribution evolves according to the partial differential equation,
\begin{equation} \label{eq:FPC}
\frac{\partial \rho}{\partial t} = -A\rho-Ax_t \frac{\partial \rho}{\partial x_t} + \frac{1}{2}BB^T \frac{\partial^2\rho}{\partial x^2_t}-\rho \int_M \lambda(x_t,m) dm,
\end{equation}
where $M$ is the full space of possible waveform features (marks) over which the joint mark intensity model is defined.

In order to define the prior distribution for each spike, we set the initial condition $\rho(t_{\mathrm{prev}},x_{t_{\mathrm{prev}}}) = f_{\mathrm{post}}(x_{t_{\mathrm{prev}}})$ equal to the posterior distribution after observing the previous spike and solve Eq. \ref{eq:FPD} or \ref{eq:FPC} (depending on whether we use clustered or sorted spike-based models) for the time of the current spike. We then set $f_{\mathrm{prior}}(x_{t_{\mathrm{current}}}) = \rho(t_{\mathrm{current}},x_{t_{\mathrm{current}}})$, where $t_{\mathrm{current}}$ is the time of the spike whose information content we are assessing. This prior distribution therefore contains all of the information up to the previous spike, and the additional information of not observing another spike since that time, until the currently observed spike. 

We call the information obtained from a spike based on the uninformative prior the \emph{isolated information} from that spike, and the information obtained based on the point process filter prior the \emph{incremental information} from that spike. Our goal is to compare the isolated and incremental information from each spike for both the spike sorted and clusterless models of neural spiking.

\subsection{Hippocampal data and modeling details}
We examined the information that hippocampal place cells contain about spatial location in a memory-guided spatial navigation task in rats. The data consist of electrophysiological recordings of tetrodes implanted in the CA1 and CA3 hippocampal areas of two male Long-Evan rats trained to perform an alternation task in a W-shaped maze \cite{Karlsson2015}. Electrophysiological data were collected from a total of 10 tetrodes, distributed bilaterally and centered at the anterior-posterior (AP) coordinates of 3.7 mm and the medial-lateral (ML) coordinates of 3.7 mm. These tetrodes were strategically placed within oval cannulas to effectively target the CA1 and CA3 regions of the hippocampus, as described in detail by Frank et al. \cite{frank2004}. The data of the sorted spikes used in the decoding analysis were obtained from neurons identified and sorted using custom software (MatClust, M. Karlsson). Two variables, peak amplitude and width of spike waveform, were used to group spikes and identify individual neurons. Neurons with low amplitude or noise were disregarded by the sorting algorithm \cite{karlsson2008}.

Previous analyses of this dataset have shown improved decoding from clusterless, population-level spiking models over spike sorted models of individual place fields, but the reason for this improvement was unexplored. For this paper, we focus on five distinct epochs, utilizing 10 tetrodes for each rat. Both spike sorting and our clusterless models were based on the peak amplitude of the triggered spike waveform from each of the four tetrode channels.

The rat's location was recorded using video imaging with a resolution of 30 Hz. We linearized the W-shaped maze so that a value of 0 represents the tip of the center arm, negative values represent distance from the center arm on leftward trajectories, and positive values represent distance from the center arm on rightward trajectories. We let $x_t$ represent the location of the rat at time $t$, based on this linearization. Our goal is to assess the amount of information available in each observed spike on the location of the rat. 

For spike sorted data, we fit receptive field models for each neuron using a non-parametric kernel-based model. Given a training set with position data recorded in discrete time steps $T = \{t_1, \cdot,t_N\}$, during which the neuron $j$ has spikes at times $S^j = \{s^j_1, \cdot, s^j_{n_j}\}$, we construct a place field model for neuron $j$ as an inhomogeneous Poisson model with rate function:
\begin{equation} \label{eq:CI}
\Lambda^j_{\mathbf{m}}(x) = \frac{\sum_{i \in S^j} K(x; x_{s^j_i}, b_x)}{\sum_{k \in T} K(x; x_{t_k}, b_x)}
\end{equation}

Here, $K(x;X,b)$ represents a one-dimensional Gaussian kernel function of the variable $x$, centered at $X$ and with a standard deviation parameter $b$. The numerator of Eq. \ref{eq:CI} is therefore the sum of Gaussian bumps centered at each observed spike location, and the denominator is the sum of bumps at each location the rat visited. The model form is analogous to an occupancy-normalized histogram model but uses Gaussian kernels instead of the histogram. We set \( b_x \) at 5 cm (compared to a track length of 96 cm) since this value minimized the decoding error \cite{Denovellis2021, Kloosterman2013}.

For the clusterless model, we fit a joint mark intensity function that defines the probability of seeing a spike with specific waveform features at each time as a function of the rat's location. Like the place field models for sorted data, we construct a nonparametric model structure that places a Gaussian bump at the location of each spike in the training set, but now that location is in the joint space of the rat's position and the spike waveform features. For each spike, we define a $d$ dimensional mark, $\mathbf{m}$, which contains the peak height of the spike on each of the four recording electrodes of the tetrode on which it was recorded. For a training set with spikes at times $S = {s_1,...,s_n}$ and corresponding marks $M = {\mathbf{m_1},...,\mathbf{m_n}}$, we define the joint mark intensity function
\begin{equation} \label{eq:JMI}
\lambda(x, \mathbf{m}) = \frac{\sum_{i \in S} \mathcal{K}(x, \mathbf{m}; x_{s_i}, \mathbf{m_i}, bx, B_m)}{\sum_{k \in T} K(x; x_{t_k}, b_x)}.
\end{equation}
Here, $\mathcal{K}(x,\mathbf{m};X,\mathbf{m_i},b,B)$ represents a Gaussian kernel function of the $(d+1)$ dimensional of variables $x$ and $\mathbf{m}$, centered on $X$ and $\mathbf{m_i}$, and with bandwidth parameters $b$ and $B$ in the location and mark dimensions, respectively. $K(x;X,b)$ is that same one-dimensional Gaussian kernel function used for the conditional intensity model in Eq.\ref{eq:CI}. This model places Gaussian bumps in the location $\times$ mark space where the spikes in the training set actually occurred and normalizes by the occupancy. To obtain the likelihood of an observed spike as a function of position, we insert the waveform mark of observed spikes, $\mathbf{m}$ into this function, and consider the resulting function over all values of $x$. 

We explored a range of bandwidths in mark space and location and selected those that minimized the decoding error. Specifically, we chose \( B_m = cI \) for the bandwidth parameters in the mark space, where \( c = 50\mu V \) and \( I \) is the identity matrix. To compute the incremental information provided by each spike in our state-space decoding analysis, we set $A$ in Eq.\ref{eq:SS} to be the identity matrix so that the state update model represented a random walk. $B$ in Eq.\ref{eq:SS} was chosen such that the rat's fastest movement speed when running back and forth along the track corresponds to one standard deviation for each discrete time step of the random walk, equal to 1 cm over each 33 ms time step. Eqs. \ref{eq:FPD} and \ref{eq:FPC} were solved using the Euler method with a discrete time step of 33 ms. 

Our analyses were based on a complete leave-one-out cross-validation approach. For each spike, we assessed its information based on the spiking models in Eqs. \ref{eq:CI} and \ref{eq:JMI} using a training set that included every spike in the data except the one being evaluated.

\section{Results}

To provide intuition about these measures, we first illustrate the decoding results for spiking activity recorded by a single tetrode in the CA1 region of the rat hippocampus during a 10-minute experiment recorded on a single day. For clarity, Figure~\ref{fig:decoding} focuses on the time interval between 4650~s and 4890~s. The figure shows the decoding results using both the full spiking data and a subset of spikes using MatClust \cite{karlsson2008}. Figure~\ref{fig:decoding}A show the decoding results for the spike-sorted model, using only the data that could be sorted with a high degree of confidence. Figure~\ref{fig:decoding}B, shows the decoding results using the clusterless model, applied to the same spikes used for sorted decoding. Figure~\ref{fig:decoding}C shows decoding results from the clusterless model using the full spiking activity, including spikes that could not be sorted with confidence. Figure~\ref{fig:decoding}D summarizes the average decoding performance across all the tetrodes recorded on various experimental days.

\begin{figure}[!h]
    \centering
    \includegraphics[width=0.9\textwidth, keepaspectratio]{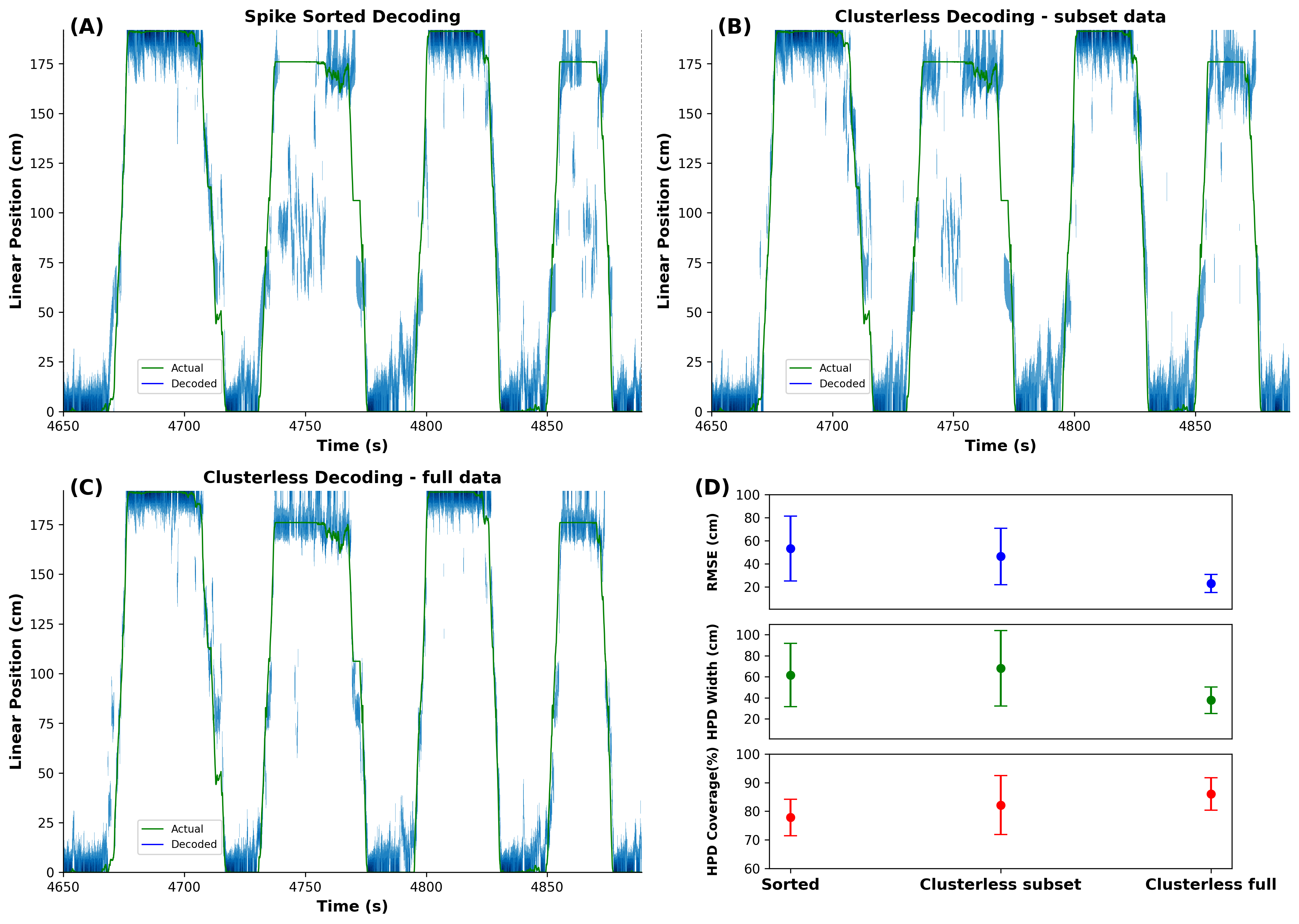}
    \caption{Decoding results from (A) Spike-sorted decoding and (B) Clusterless decoding using only spikes that could be sorted with a high degree of confidence, (C) Clusterless decoding using all triggered spikes. The green line represents the actual rat's position, and the blue region shows the posterior density at each time step. (D) Average decoding performance across different tetrodes from multiple experimental periods.}
    \label{fig:decoding}
\end{figure}

Figure~\ref{fig:decoding}A–B use the same subset of spikes, excluding those that could not be reliably sorted, while Figure~\ref{fig:decoding}C uses the full spiking activity. Consistent with previous findings \cite{Deng2015}, the decoding accuracy is higher with the clusterless model using the full spiking activity, showing reduced rMSE (clusterless: 15.93 vs. sorted: 31.97), smaller HPD regions (clusterless: 20.61 vs. sorted: 44.85), and better HPD coverage (clusterless: \SI{88}{\percent} vs. sorted: \SI{83}{\percent}) of the true position. This improvement is particularly evident during the period from 4710 s to 4790 s, when the sorted decoder consistently underestimates the true position. Although there is a slight improvement when using the clusterless decoding with subset of data Figure~\ref{fig:decoding}B (rMSE: 29.61, HPD regions: 39.51, HPD coverage: \SI{87}{\percent}), the results are significantly better with the clusterless model using the full dataset in Figure~\ref{fig:decoding}C.

Figure~\ref{fig:decoding}D shows the average decoding performance,  the width of a 0.95 highest posterior density (HPD) region, and its coverage of the rat's true location, computed across all of the tetrodes. The rMSE of the estimates decreases from the sorted decoder to the clusterless decoder. The highest posterior density (HPD) width shows a slight improvement with clusterless decoding using subset of data and a significant improvement with the clusterless model using the full spiking data. Additionally, the average HPD coverage percentage is higher with clusterless decoding both with subset data and with full spiking activity compared to spike-sorted decoding.

In prior research, it was unclear whether the observed improvement in decoding was primarily due to additional spikes excluded during spike sorting or to the enhanced information captured in individual spikes by the clusterless model. The fact that the decoding results in Figure~\ref{fig:decoding}A-B use the same spikes suggests that the improvement arises at least in part from the clusterless model's ability to extract more information from each individual spike or to make better use of the available data. To investigate this further, we measured various information metrics across multiple tetrodes using both isolated and incremental information.

\begin{figure}[!h]
    \centering
    \includegraphics[width=0.95\textwidth, keepaspectratio]{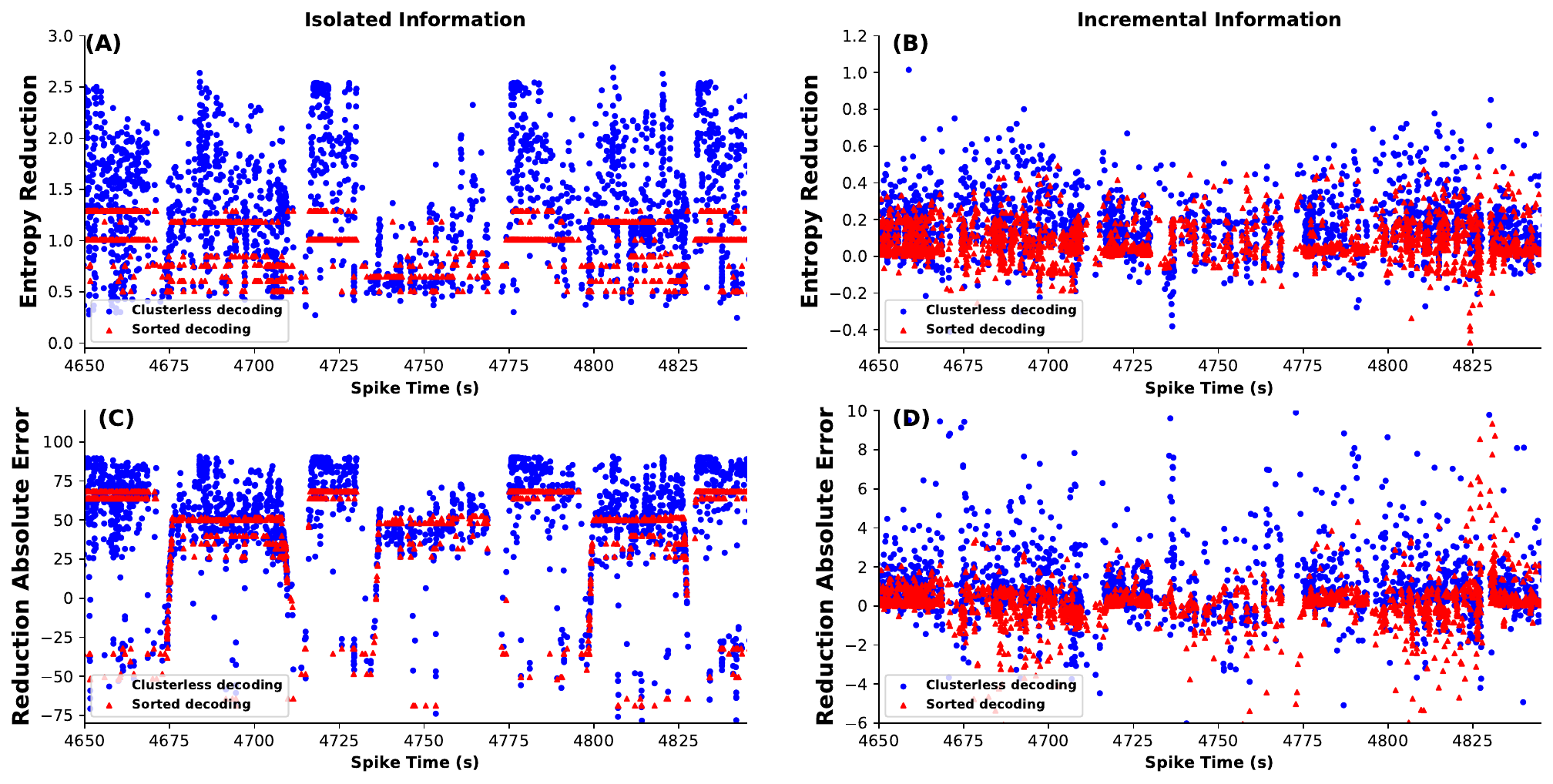}
    \caption{Top panels show the reduction in entropy, while bottom panels show the reduction in absolute error for each spike as functions of time for both spike-sorted decoding (red points) and clusterless decoding (blue points). Each decoding method is evaluated using both isolated information (left panels) and incremental information (right panels). Note that both the spike-sorted and clusterless models use the same subset of spikes.}
    \label{fig:entropy}
\end{figure}

Figure~\ref{fig:entropy} compares the reduction in entropy (RE), and the reduction in absolute error (RAEr) between the prior and posterior models. Figure~\ref{fig:entropy}A shows the reduction in entropy of a uniform prior (isolated information) using spike-sorted (red) and clusterless (blue) decoding. Since spikes sorted into the same cluster have the same intensity, they provide the same reduction in entropy relative to the uniform distribution, so that the red points occur at discrete levels. The same spikes from clusterless models have mark-dependent intensities, leading to a broader range of values for the reduction in entropy. Notably, we observe that the reduction in entropy from a uniform prior is consistently smaller in the spike-sorted model compared to the clusterless model. Since the uniform prior has a fixed entropy, the reduction is determined solely by the entropy of the normalized likelihood of the spike as a function of position. This means that spikes from the clusterless model produce likelihoods that are more concentrated in space and therefore more informative than the spike sorted models. 

Figure~\ref{fig:entropy}B shows the reduction in entropy for each spike based on the one-step prior distribution from the state-space decoder. This reduction in entropy represents the one-step (incremental) information provided by each spike during the ongoing decoding process. Equivalently, we can think of this entropy reduction as the additional information provided by each spike, when accounting for all of the information from previous spikes. Notably, this reduction in entropy is of much smaller magnitude than the reduction in entropy from the uniform prior, and can even take on negative values when the posterior after observing the spike has greater entropy than the one-step prediction prior. In that case, observing a spike makes the decoder less certain of the rat's position. We note that there seems to remain an improvement in the reduction in entropy due to the clusterless model spikes over the sorted spikes, but this improvement is relatively smaller and less consistent than it was when examining the isolated information. This suggests that the decoding bias induced by spike sorting is mitigated rather than exacerbated with increasing amounts of data.

Figure~\ref{fig:entropy}C illustrates the reduction in absolute error (RAEr) using isolated information for clusterless and spike-sorted decoding. The reduction in absolute error tends to be smaller for spike-sorted decoding compared to clusterless decoding, but this result is not as consistent as it was for the reduction in entropy. This suggests that in isolation, spikes from clusterless models tend to result in much more precise, but not necessarily much more accurate estimates of the rat's location.

Figure~\ref{fig:entropy}D demonstrates the RAEr based on incremental information. Unlike Figure~\ref{fig:entropy}C, the values for the spike sorted model (in red) again seem consistently smaller than the clusterless model (in blue), suggesting that aggregating information from spikes through time tends to reduce the error of clusterless decoding more than from spike-sorted decoding.

\begin{figure}[!h]
    \centering 
    \includegraphics[width=0.95\textwidth, keepaspectratio]{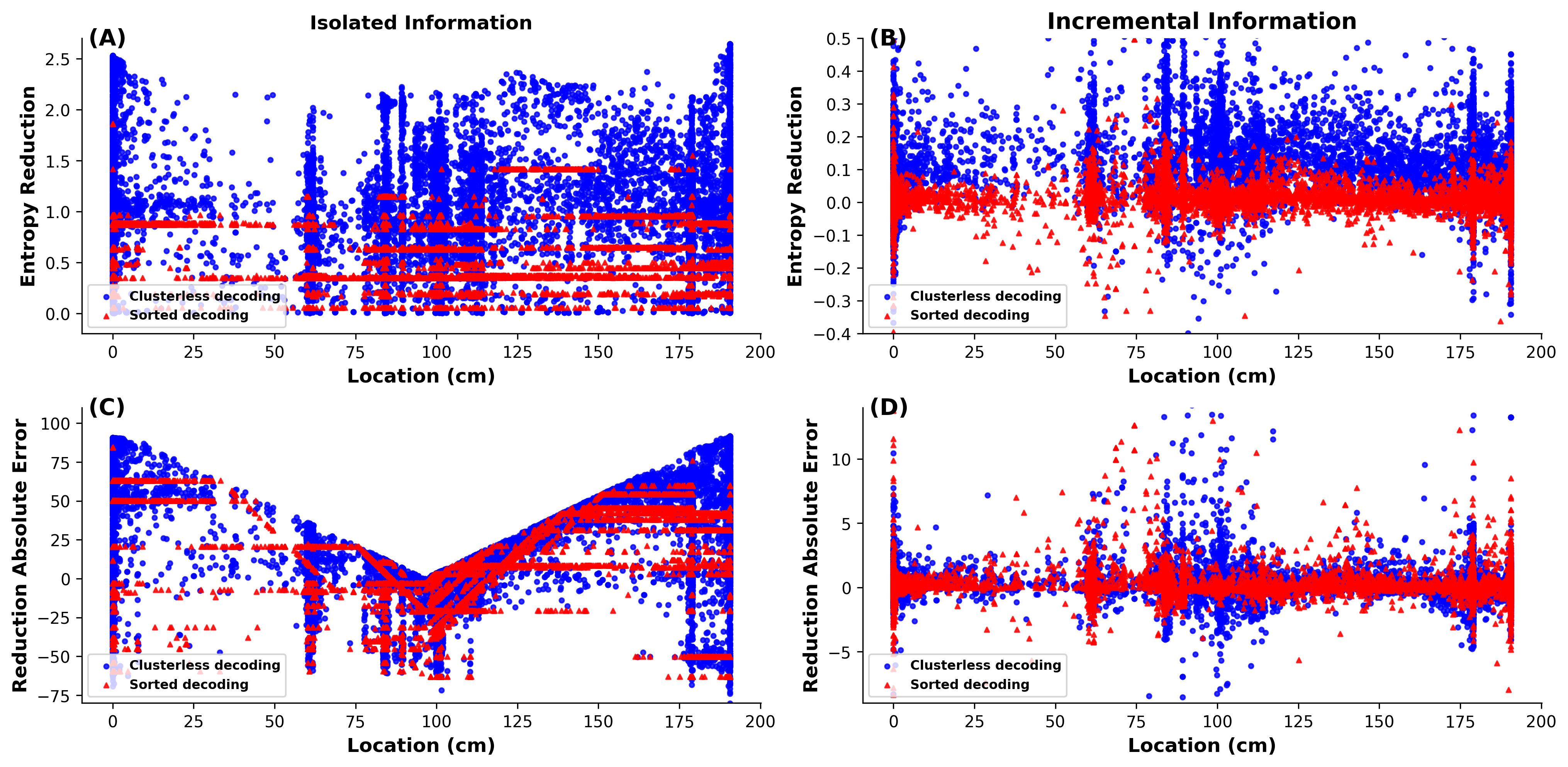}
    \caption{
Reduction in entropy and absolute error versus the rat's position at spike times. 
Panels (A) and (B) show the reduction in entropy, while panels (C) and (D) show the reduction in absolute error. Left panels correspond to isolated information while right panels correspond to incremental information. Red points represent spike-sorted decoding, and blue points represent clusterless decoding.}
    \label{fig:location}
\end{figure}

Another question we address is whether the reduction in entropy and absolute error differ at specific locations along the track. Figure~\ref{fig:location}A-B show the reduction in entropy, while Figures \ref{fig:location}C-D depict the reduction in absolute error for each spike as a function of the position of the rat at the time of the spike. The left panels represent the isolated information, and the right panels show the reduction relative to the decoded one-step prediction distribution (incremental information) using prior spikes. In both plots, red dots indicate the reduction in information due to sorted spikes, and blue dots represent the reduction of the clusterless model. Once again we observe that the reductions in entropy and error tend to be larger for the clusterless models. 

In Figure~\ref{fig:location}A we see clear horizontal red streaks corresponding to individual neurons that extend across most of the environment, suggesting that some of the sorted neurons have fairly broad tuning, which explains the relatively lower reduction in entropy values. For the clusterless model in blue, we see more vertical blobs, suggesting that at many locations we observe spikes whose waveform features vary, some of which provide relatively precise estimates of location, some of which provide relatively broad estimates, and everything in between. We observe similar vertical blobs when considering the incremental information from the clusterless model in Figure~\ref{fig:location}B.  

Figure~\ref{fig:location}C shows the reduction in absolute error from a uniform prior (with mean estimate in the center of the track) based on each spike using the spike-sorted (red) or clusterless (blue) models. The shape of the upper bound on these points reflects the maximum improvement of the estimate of the rat's location from the prior estimate at the center of the track to the true location of the rat. The fact that so many of the blue points cluster near these bounds suggests that the joint mark intensity tends to provide unbiased estimates of the rat's position on the track. However, there is notable departure from these bounds, particularly at the endpoints of the track. One possible explanation is that as the rat slows down at the endpoints, spiking activity shifts away from encoding its current location and instead reflects non-local phenomena such as replay \cite{wilson1994reactivation,karlsson2009awake, davidson2009}. This difference at the endpoints is substantially mitigated when estimating the decoding information from all previous spiking (Figure~\ref{fig:location}D).

\begin{figure}[!h]
    \centering
    \includegraphics[width=0.95\textwidth, keepaspectratio]{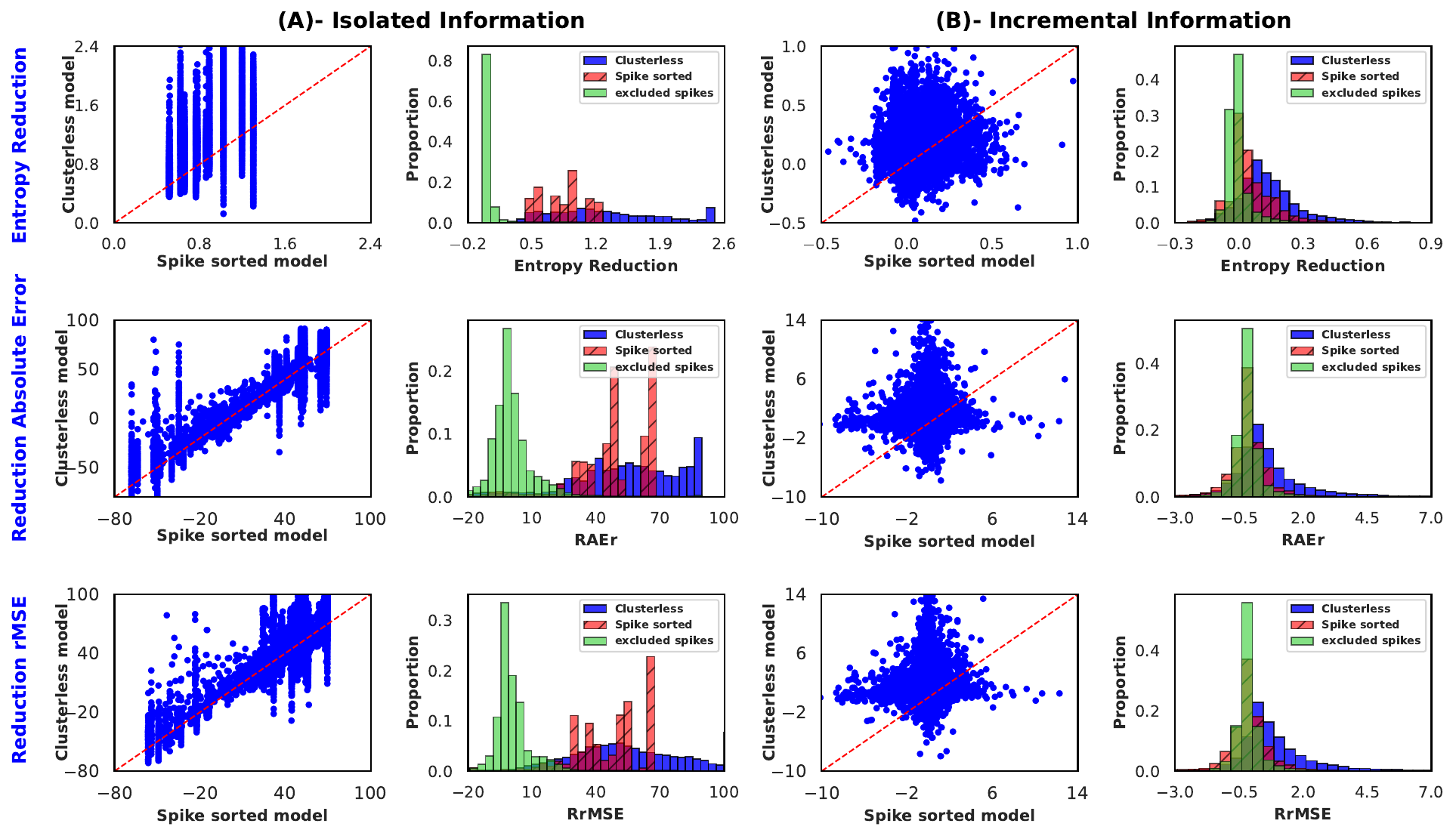}
    \caption{Reduction in entropy (first row), reduction in absolute error (RAEr) (second row), and reduction in root mean square error (RrMSE) (third row) for isolated (A) and incremental information (B). Each scatterplot shows spike-by-spike comparisons of information measures between spike-sorted and clusterless decoding for the subset of spikes that were sortable; each histogram shows the distributions of these measures for the spike-sorted model (red), the clusterless model applied to the sorted spikes (blue), and the clusterless model applied to the spikes excluded by spike sorting (green).}
    \label{fig:association}
\end{figure}

Previous results show that clusterless decoding provides more information than spike-sorted decoding. To assess the degree to which the information from individual spikes contributes to this improvement, we compared the isolated and incremental information between these models for individual spikes, illustrated in Figure~\ref{fig:association}. Panels (A) and (B) present results for isolated and incremental information, respectively. In each scatterplot, the y-axis represents the value of each information measure for the clusterless model, while the x-axis represents the value for the spike-sorted model. 

Figure~\ref{fig:association}, first row, compares the reduction in entropy between clusterless and spike-sorted models on a spike-by-spike basis. Includes isolated information using the uniform prior and incremental information using the one-step prediction prior, along with their respective distributions.
For isolated information, we observe that entropy reduction in spike-sorted models occurs as fixed values for each sorted neuron, which appear as vertical lines. Many of these lines are almost entirely above the red $45^{\circ}$ line, indicating that most spikes provide more information when using the clusterless model. In general, \SI{89}{\percent} of the spikes result in a greater reduction in entropy from a uniform prior when using the clusterless model compared to the spike-sorted model. Additionally, the correlation coefficient between entropy reductions in the two models is 0.56, suggesting that the most informative sorted neurons (as the same spike data are used in both models) also tend to produce the most informative spikes in the clusterless model.
Panel (B), where incremental information is used, shows a substantial reduction in the magnitude of this difference. As previously observed, the overall entropy reduction is much smaller, and some values are now negative. Despite this, \SI{76}{\percent} of the points remain above the $45^{\circ}$ line, indicating that spikes in the clusterless model are, on average, more informative than those in the spike-sorted model when accounting for the prior spiking history. However, the correlation coefficient drops to 0.27 when using the one-step prediction prior (incremental information), suggesting that the relationship between the two models weakens.
Incremental information is consistently lower than isolated information, and the correlation between spikes is much weaker in the incremental case. Spikes that are most informative in isolation are not necessarily as informative incrementally, indicating that the most informative spikes in the spike-sorted model no longer tend to be the most informative in the clusterless model. This discrepancy likely arises from substantial differences in how prior spiking information is utilized in the two models.

In Figure~\ref{fig:association}, the second row shows the reduction in absolute error, which measures the improvement in estimates under the prior and posterior distributions due to each observed spike. The comparison is made on a spike-by-spike basis between clusterless and spike-sorted decoding for both isolated and incremental information.
Most points lie above the $45^{\circ}$ line, indicating that the reduction in absolute error is greater in the clusterless model than in the spike-sorted model. In the isolated information plot,  \SI{73}{\percent} of the data points fall above the line, demonstrating that for most spikes, the reduction in absolute error is larger when using the clusterless model. When incorporating past spike information into the incremental information plot, the ratio remains similar, with approximately \SI{78}{\percent}of data points above the line. However, the reductions are more concentrated. Similar results are observed for the reduction root mean square error (RrMSE), as shown in the third row of Figure~\ref{fig:association}.

In Figure~\ref{fig:association}, the green distribution represents the reduction in the information measures of the spikes that were excluded during the data sorting process. These spikes are present in the full dataset. In the isolated model, most of these spikes contain little to no information, making them non-informative. This aligns with the reasoning behind spike sorting and the exclusion of certain spikes in many studies.
However, in the incremental model at the population level, we observe that these excluded spikes contribute some information, which is not necessarily close to zero. The proportion of entropy reduction contributed by the excluded spikes, relative to the total entropy reduction across all spikes, is \SI{48}{\percent}. Similarly, this ratio is \SI{37}{\percent} for the reduction in absolute error and \SI{43}{\percent} for the reduction in rMSE in incremental information. Since this ratio is positive and above zero, it indicates that a considerable amount of information is retained in the excluded spikes. This suggests that while spike sorting removes certain spikes, using the full spiking dataset in clusterless decoding allows us to recover meaningful information that would otherwise be discarded.

By comparing the results of isolated and incremental information, we observe that spikes that are most informative in isolation may not informative when considering in incremental information. For the same spike, reductions in entropy, absolute error, and RMSE indicate that clusterless decoding provides more informative and significant results compared to spike-sorted decoding. This highlights the need for further investigation into whether the marked point process captures additional information from spike waveform characteristics and how these measures relate to spike waveforms.

\begin{figure}[!h]  
    \centering
    \includegraphics[width=0.9\textwidth, keepaspectratio]{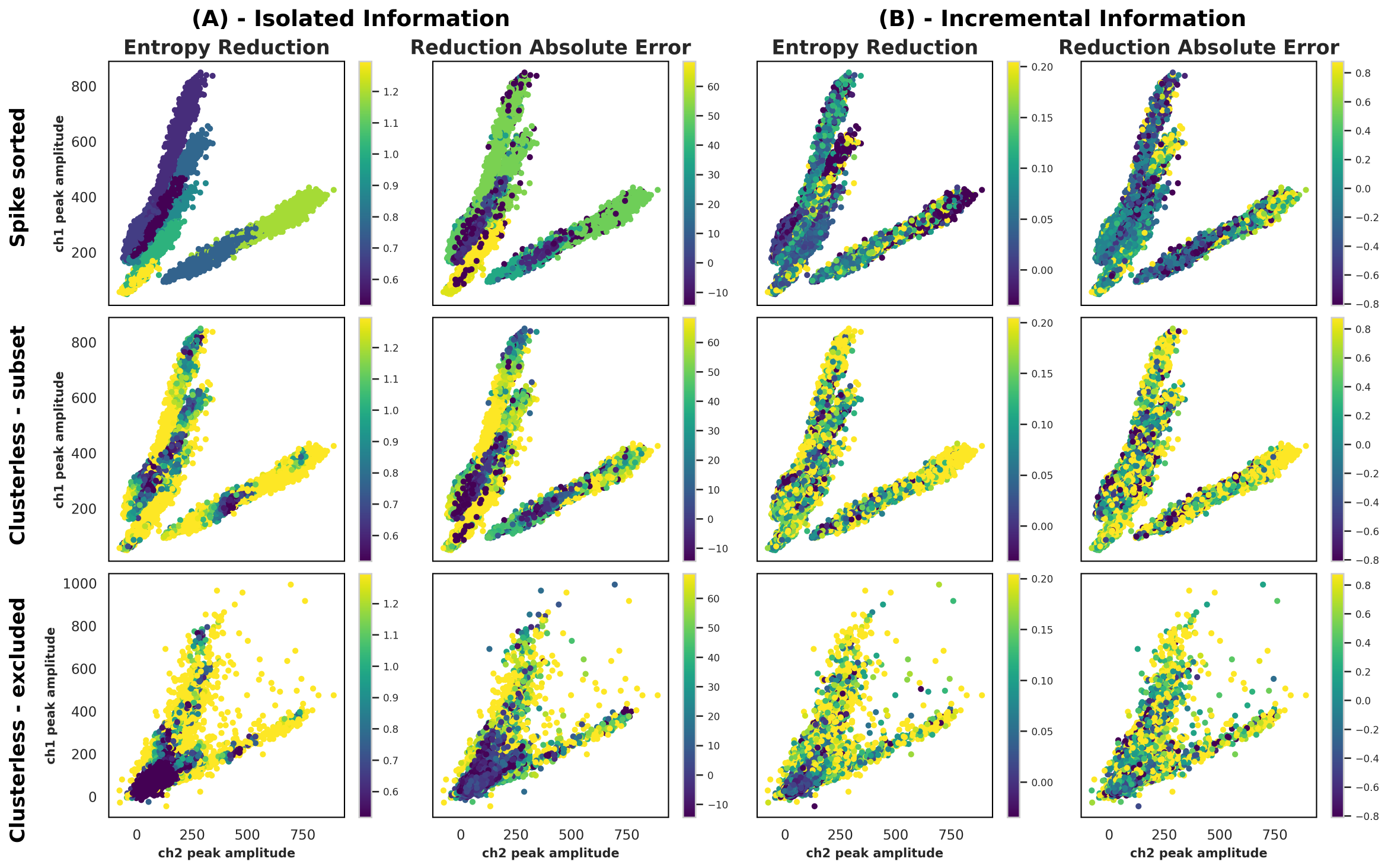} 
    \caption{Reduction in entropy and absolute error for spikes recorded from a single tetrode in the CA1 region of the hippocampus. The x- and y-axes represent spike amplitudes recorded on the $1^{st}$ and $2^{nd}$ tetrode channels (in $\mu V$). The figure compares spike-sorted and clusterless models, including results for spikes excluded during spike sorting. The color scale indicates the magnitude of entropy and absolute error reduction.}
    \label{fig:mark}  
\end{figure}

Figure \ref{fig:mark} shows the reduction in entropy and absolute error for spikes recorded from a single tetrode in the CA1 region of the hippocampus. The x- and y-axes represent the peak amplitudes of each spike recorded on the first and second channels of the tetrode, respectively. The figure compares the same spikes across the spike-sorted and clusterless models that have been sorted, as well as the clusterless model results for spikes excluded during spike sorting. The color represents the magnitude of entropy and absolute error reduction. This comparison helps assess how clusterless decoding captures additional information from both included and excluded spikes, providing insight into the effectiveness of these models.

The first and second rows of the figure \ref{fig:mark} use the same subset of spikes that have been sorted: The first row presents results from spike-sorted decoding, while the second row shows results from clusterless decoding applied to the same subset spikes. In contrast, the third row contains a different set of spikes, specifically those excluded during spike sorting. These spikes do not overlap with those in the first two rows, but, as in the second row, their decoding results are obtained using clusterless marked point process decoding. The color represents the magnitude of the reduction in entropy and absolute error.

Figures \ref{fig:mark}A (left panel) show the reduction in entropy and absolute error calculated relative to a uniform prior, referred to as isolated information. Figure\ref{fig:mark}B (right panel) presents the same measures using incremental information, which incorporates one-step information from previous spiking history. As expected, each spike-defined cluster has its own receptive field model, providing a consistent amount of information when considered independently. For example, in the spike-sorted model, the reduction in entropy relative to a uniform prior ranges from approximately 0.5 to 1.4 (Figure \ref{fig:entropy}A). To improve clarity, the information measures in all subplots are plotted over smaller intervals to enhance visual resolution.

The second row of Figure \ref{fig:mark}A shows the results of clusterless decoding, analyzing the same spiking data as in the first row. This panel illustrates both the entropy reduction and the reduction in absolute error for spikes using uniform prior in the clusterless model. Unlike the spike-sorted model, where the information associated with each spike remains consistent across clusters, the clusterless model shows variability. However, the information per spike is generally higher in the clusterless model (yellow color). The reduction in entropy relative to a uniform prior for the clusterless model ranges from 0.33 to 2.6 ( Figure \ref{fig:entropy}A ). Comparing the clusterless results with those of the spike-sorted model in the first row, we observe a higher reduction in entropy for individual spikes across both low- and high-amplitude spikes (marks or channels). A similar pattern of greater reduction is evident for the absolute error, demonstrating consistent improvements in isolated information between the sorted and clusterless models.

The third row of Figure \ref{fig:mark}A  shows  the reduction in entropy and the absolute error for the clusterless model, focusing on spikes excluded from the spike-sorting model (unclustered spikes). Not surprisingly, this adds a dense region of relatively low information spikes with amplitudes less than 200 $\mu$V and  close to 0 also regions with high information (yellow color) with both low-amplitude spikes and high-amplitude spikes. We classified low amplitude spikes as the lowest 25th percentile of recorded spike waveforms across the marks (peak amplitude in 4 channels), and high amplitude spikes as the top 75th percentile of recorded spike waveforms. In fact,  more than \SI{54}{\percent} of the recorded spikes are low-amplitude spikes, \SI{41}{\percent} are high amplitude spikes. Of these \SI{99}{\percent} of low-amplitude spikes and \SI{75}{\percent} of high amplitude spikes, they are discarded in the spike sorted model. Interestingly, these low-amplitude spikes still provide substantial information. 

Additionally, many high-amplitude spikes are visible in this row that do not appear in the upper rows because they could not be reliably sorted. Nearly all of these spikes carry substantial information and are effectively used by the clusterless model to improve decoding accuracy. This distinction is clearly illustrated in the third row of Figure \ref{fig:mark}A, which shows the isolated contributions to entropy reduction and absolute error.

Figures \ref{fig:mark}B show the one-step reduction in entropy and absolute error (incremental information) which considers the information from all previous spikes. Note that the color scale for the right panels ranges over values much smaller than those for the left panels because much of the potential information available in each spike is already accounted for in the decoded estimate from previous spikes.  The top, middle, and bottom plots represent the reduction in entropy of the sorted model, the clusterless model that uses only subset spikes, and the clusterless model that uses excluded spikes, respectively. Unlike the plots on the left, these plots of incremental information during decoding show much more heterogeneous amounts of entropy reduction within each visible cluster. This suggests that the amount of information that comes from each spike is less a function of the coding properties of the particular neuron, and more a function of the state of knowledge in the decoder at the time the neuron is fired. Using the same subset of spikes, the spike-sorted incremental model shows a smaller reduction in entropy and absolute error than the clusterless incremental model. In panel (B), this appears as more uniformly deep-blue maps (smaller reduction) in the incremental model using sorted decoding versus more yellow regions (larger reduction) in the second row using clusterless decoding in incremental model. 

In particular, in Figure \ref{fig:mark}B, third row, we observe that among the spikes typically discarded during spike sorting, some provide medium (green points) and large (yellow points) reductions in entropy and absolute error. For lower-amplitude spikes, the distinction between the one-step information available from low-amplitude spikes and sortable higher-amplitude spikes becomes less clear. Additionally, a number of high-amplitude spikes are discarded during spike sorting, and these tend to provide more information than the low-amplitude spikes that are discarded. Observable points (yellow) still exhibit high reductions in entropy and absolute error. 

This comparison helps assess how clusterless decoding captures additional information from both included and excluded spikes. The difference between isolated and incremental information highlights that excluded spikes still carry valuable information, and incremental information allows us to analyze past spiking history more effectively. In Figure \ref{fig:mark}-third row, we see this distinction between isolated and incremental information.  Analyzing the excluded spikes in isolation (no spiking history; uniform prior) shows they are less informative than in the incremental model. In the isolated analysis, entropy reduction is amplitude dependent: a large deep-blue region below $200\,\mu\mathrm{V}$ indicates little contribution from low-amplitude spikes. In contrast, the incremental model shows no simple relationship between amplitude and information. Some low-amplitude spikes remain highly informative, while not all large spikes contribute equally. Even spikes typically discarded during sorting can carry meaningful information about position. These excluded spikes remain part of the full spiking data used for clusterless decoding. Our goal is to evaluate the impact of these unsortable spikes by quantifying the information lost when they are excluded from the spike-sorted dataset. Their presence and contribution to the overall information content suggest that they may contain valuable and usable information for decoding. Motivated by these findings, we next focus on information measures derived from incremental models in our analysis.

As shown in Figure~\ref{fig:association}, clusterless decoding produces lower entropy, absolute error, and RMSE than spike-sorted decoding when both are applied to the same dataset. Performance improves further when the clusterless model uses the full spike train, incorporating not only the spikes assigned to clusters but also those excluded during sorting. To quantify the information contained in the complete spiking activity, we define cumulative information as the total entropy reduction contributed by all detected spikes, both sorted and unsorted.

To illustrate how cumulative information differs between the two approaches, we select a short time bin during active running that contains the spikes present in the sorted dataset as well as all spikes in the full recording, including those excluded during spike sorting.
Within each interval, the clusterless model accumulates the prior entropy reductions from these additional unsortable spikes, adds the contribution of the next sorted spike, and compares this total with the spike-sorted result containing only the sorted spike. A positive cumulative difference indicates that the excluded spikes contribute additional information. After each sorted spike, the cumulative sum is reset to isolate the information contributed by the next set of unsorted spikes.

\begin{figure}[!h]
  \centering
  \includegraphics[width=\columnwidth, max height=\textheight, keepaspectratio]{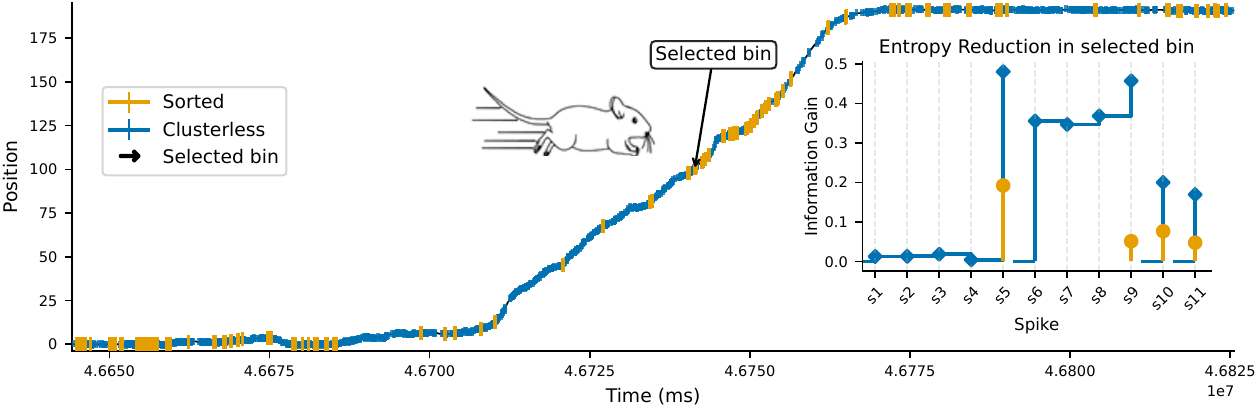}

  \caption{Rat trajectory with a specific selected time bin during rapid running. The blue vertical tick marks indicate spikes from the full (clusterless) dataset, and yellow tick marks indicate spikes retained after spike sorting. The panel to the right shows per-spike entropy reduction (ER) within the selected bin: blue traces correspond to cumulative ER for clusterless decoding, and yellow markers indicate ER for spike-sorted decoding.}
  \label{fig:selected-bins}
\end{figure}

Figure~\ref{fig:selected-bins} shows a short segment of the rat’s trajectory containing the selected time bin, which includes four sorted spikes and eleven total detected spikes. Blue markers denote all detected spikes from the full dataset, including those shared with the sorted data and the additional spikes excluded by spike sorting. Within this interval, we compare the cumulative information from the clusterless model (blue trace), which uses all detected spikes, and the spike-sorted model (yellow dots), which includes only the four sorted spikes.

The first four spikes appear only in the clusterless dataset and each produces a small reduction in entropy, reflecting only minor information gain. The fifth spike (s5), the first shared between the two models, shows a greater entropy reduction in the clusterless model. The additional spikes unique to the clusterless model do not create large cumulative information gains from prior spiking history; rather, the information improvement arises from reduced bias in decoding from s5. 

The sixth spike (s6) in the clusterless model produces a substantial increase in information gain, which is propagated forward. The next sortable spike is the ninth spike (s9), and if the information in s6 and s9 were redundant, we might expect the entropy reduction from s9 for the spike sorted model to be much larger than for the clusterless model since the clusterless model already accounted for that information from s6. Instead, we see that the entropy reduction from s9 is similar for the spike sorted and clusterless models. In this example, the clusterless model achieves higher information gain because it integrates information from previous spikes that were unavailable to the spike sorted decoder.

The final two spikes are common to both models, and the clusterless decoder again yields a substantially higher information gain, even in the absence of prior information. This finding indicates that, independent of cumulative effects, the marked point process framework captures additional information from spike waveforms and reduces bias, which contributes to the improvement observed in this case.

This example highlights two ways in which clusterless decoding improves upon spike sorting. First, spikes removed by sorting provide non\mbox{-}redundant information that accumulates and enhances the contribution of later common spikes. Second, some individual unsorted spikes themselves carry more information than their sorted counterparts. This additional contribution arises because the clusterless model directly uses waveform features from all spikes, including those excluded by sorting, allowing them to convey non\mbox{-}redundant information about position. 

\begin{figure}[!h]
  \centering
\includegraphics[width=\columnwidth, keepaspectratio]{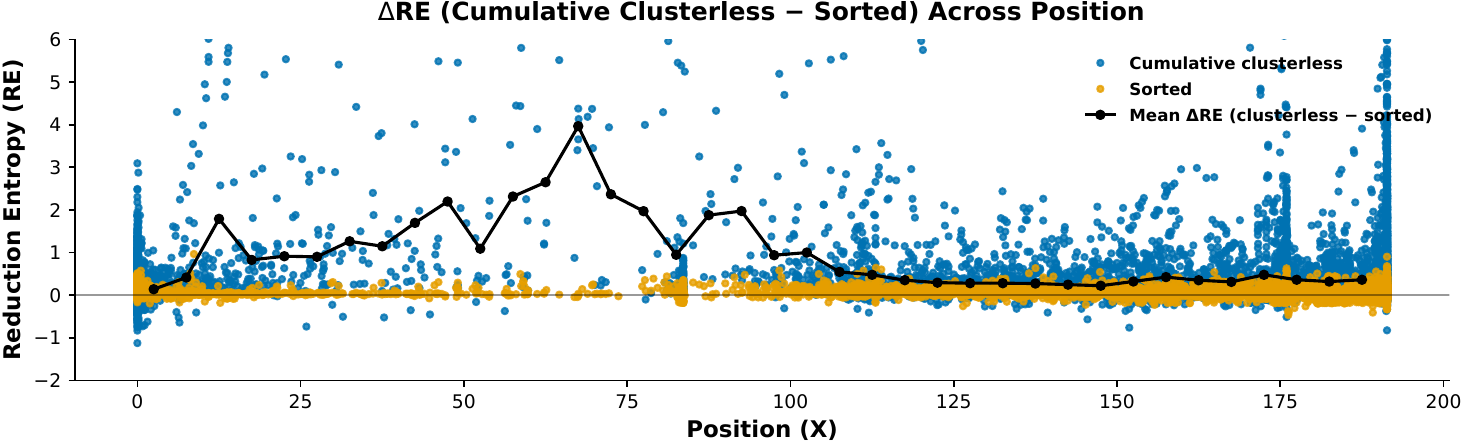}
  \caption{
Comparison of incremental information (entropy reduction) between clusterless and spike-sorted decoding. The figure illustrates the effect of additional spikes and the cumulative information captured by the clusterless model (blue points) compared with the spike-sorted model (yellow points), which includes only sorted spikes. The black line shows the mean difference in information across position bins.
}
  \label{fig:overal-bins}
\end{figure}

Figure~\ref{fig:overal-bins} shows the cumulative reduction in entropy obtained from the clusterless model applied to the full spiking data, highlighting the additional information retained when unsorted spikes are included. Blue and yellow points represent cumulative information from all spikes and from sorted spikes, respectively, while the black line indicates the mean difference in incremental entropy reduction ($\Delta$RE) between the two models across position bins.

When the density of sortable neurons is high (positions $> 100$), both models perform similarly, though the clusterless model maintains modest improvements due to lower bias. 
In contrast, when sortable neurons are sparse (positions $\approx 10$--$100$), the cumulative clusterless model captures substantially more information because unsorted spikes provide additional contributions absent in the spike\mbox{-}sorted dataset.

These results illustrate that the benefit of clusterless decoding depends on the spatial density of sortable neurons: it is greatest where sorting coverage is limited, as incorporating all spikes compensates for the lack of identified units. 
Clusterless decoding thus captures spatial information that is lost under spike sorting. 
Spikes discarded during sorting make a meaningful positive contribution to uncertainty reduction, as observed consistently across different experiments. 
Overall, the improvement in decoding performance arises from both the inclusion of previously discarded spikes and the reduction of model bias. 
This demonstrates that the clusterless approach more effectively exploits available information by leveraging the full spike train and waveform features. 
Taken together, these results highlight the consistent advantage of clusterless decoding within individual recordings. 
To assess whether this advantage generalizes more broadly, we next examine multi\mbox{-}electrode datasets and summarize performance across experiments.

\subsection {Summary of Multi-Electrode and Cross-Experiment Results}

 In this section, we summarize the results using multiple tetrodes from two rats in multiple experimental sessions to assess the consistency and variability of our findings, focusing on whether changes in bias or spike exclusion predominated under various experimental conditions. We evaluated decoding performance and information measures across epochs, environments, and experiments. These analyses demonstrate the robustness and coherence of our findings in various experiments and datasets. Figure~\ref{fig:summary}, panel (A) summarizes the results in multiple experiments using three decoding models: (1) clusterless decoding with the full spiking dataset, (2) clusterless decoding with a subset of spiking data, and (3) spike-sorted decoding using the same subset.

\begin{figure}[!h]
    \centering 
    \includegraphics[width=0.9\textwidth]{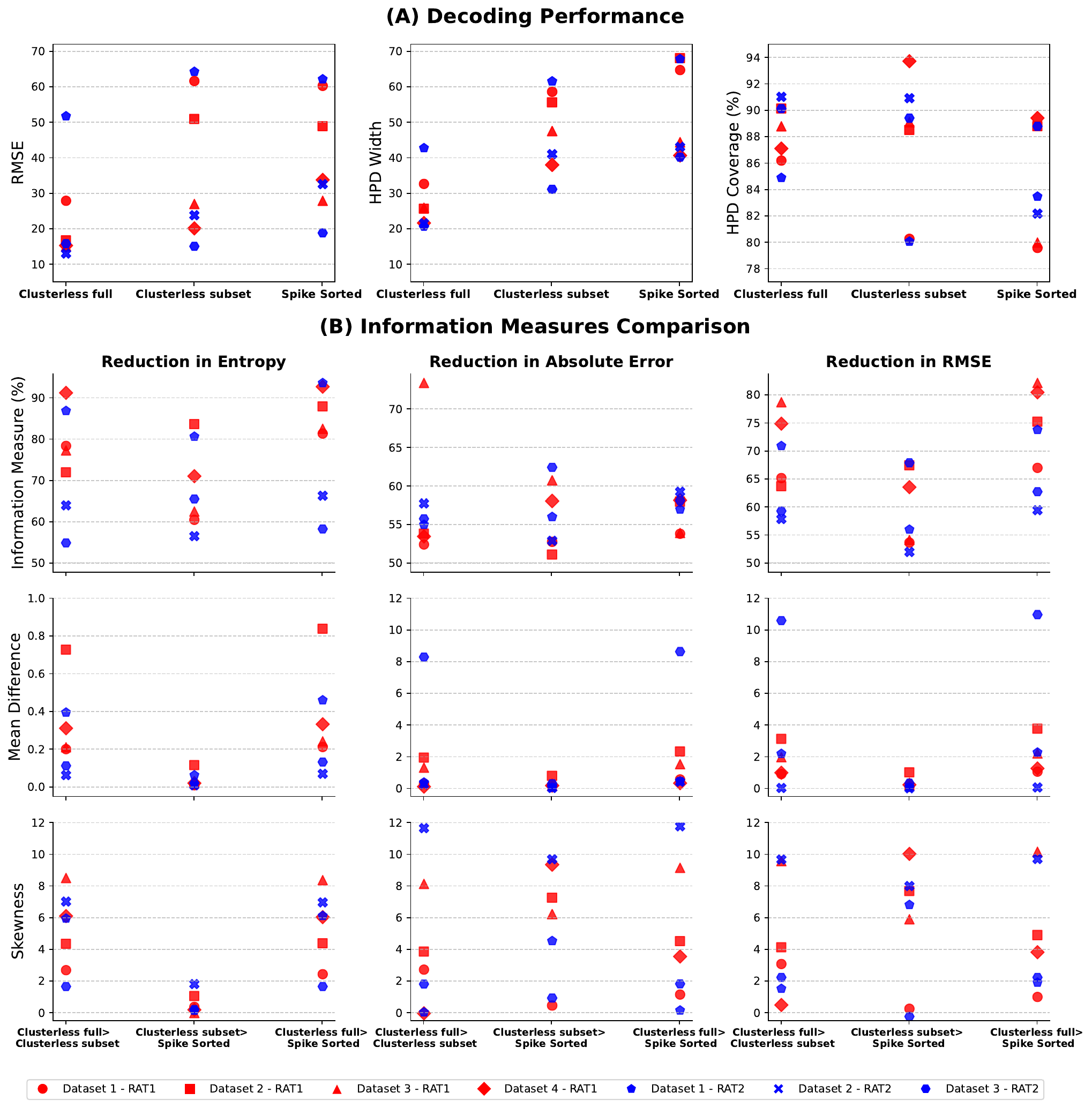}
\caption{
(A) Decoding performance across datasets from two rats, measured by RMSE, HPD width, and HPD coverage. Red and blue markers represent Rat 1 and Rat 2, respectively; marker shapes indicate individual datasets.
(B) First row: counts of time points where the clusterless full model outperforms the subset and spike-sorted models. Second row: mean difference in performance at each time point. Third row: skewness of these differences between the clusterless full model and the other two models with subset data.
}
    \label{fig:summary}
\end{figure}

The plot on the left shows the root mean square error (RMSE) between the expected (decoded) position and the actual position of the rat. Overall, the clusterless method using the full spiking data tends to produce lower RMSEs than the clusterless method using the subset, which in turn tends to outperform the spike-sorted method. However, this trend does not hold for every epoch and every rat. For example, in the red square and blue trapezoid datasets, the RMSE of the clusterless subset model slightly exceeds that of the spike-sorted model. In the red triangle dataset, the RMSEs of the clusterless subset and spike-sorted models are nearly identical.

The middle plot in panel (A) shows the width of the 98\% highest posterior density (HPD) region, defined as the number of position bins required to accumulate 98\% of the posterior probability mass. This metric reflects the average width of the decoder confidence region. On average, the HPD width is smallest for the clusterless model using the full spiking data, followed by the clusterless subset, with the spike-sorted model exhibiting the widest confidence regions. This indicates that the full clusterless model provides greater certainty in decoding the rat position across all datasets. However, this trend is not consistent in all cases for subset data. For example, in one dataset (represented by the red triangle), the HPD width for the clusterless subset is slightly larger than that of the spike-sorted model, even though their RMSEs were nearly the same.

The rightmost plot in panel (A) shows the accuracy of the highest posterior density (HPD), defined as the proportion of time steps in which the decoder successfully captured the true position within the 98\% highest posterior density region. This metric reflects how often the true position falls within the 98\% confidence region. The clusterless model using the full spiking dataset generally achieves higher and more consistent HPD coverage than the clusterless subset, which in turn outperforms the spike-sorted model. However, this pattern is more variable across datasets than what is observed for the RMSE or HPD width. For instance, in the dataset represented by the blue cross, coverage is comparable between the full and subset clusterless models, but substantially lower for the spike-sorted model. In contrast, the red rhombus dataset shows higher coverage for the clusterless subset than for the spike-sorted model, which in turn outperforms the full clusterless model. In the blue trapezoid dataset, the full clusterless model again provides the highest coverage, but the spike-sorted model notably outperforms the clusterless subset. Interestingly, in this same dataset, the RMSE for the clusterless subset is only slightly higher than that of the spike-sorted model.

These observations highlight that improvements in RMSE do not necessarily correspond to improvements in HPD coverage. This underscores the importance of evaluating multiple performance metrics when comparing decoding models. Such an analysis also motivates a deeper investigation into information-theoretic measures to better understand the uncertainty and reliability of the decoding models.

Figure~\ref{fig:summary}B compares the reduction in entropy, absolute error, and root mean square error (RMSE) across three decoding models: clusterless decoding using the full spiking dataset, clusterless decoding using the same subset of spikes as the spike-sorted model, and spike-sorted decoding. These comparisons are made at each spike time using the spikes common to both the full and subset datasets. As shown in Figure~\ref{fig:selected-bins}, the full clusterless model accumulates information not only from the current spike but also from preceding shared spikes, yielding a more comprehensive estimate at each time point.
Panel (B) also quantifies how often each model outperforms the others across the three evaluation metrics: entropy reduction (left), absolute error reduction (middle), and RMSE reduction (right). The top row shows the percentage of cases in which the full clusterless model surpasses the other two models, while the middle points on the x-axis indicate direct comparisons between the clusterless subset and spike-sorted models, which share the same spike data.
Across all metrics, values consistently above 50\% reveal a clear population-level trend: decoding performance improves from the spike-sorted model to the clusterless subset and further improves with the full clusterless model. Although the magnitude of improvement varies across datasets, the overall pattern remains consistent.

This consistent ordering suggests two distinct sources of improvement. First, the improvement of the clusterless full model over the subset model likely reflects the value of retaining additional spikes that are excluded from the subset. Second, the improvement of the clusterless subset over the spike-sorted model suggests a reduction in bias introduced by the spike-sorting process. Since both models use the same subset of spikes, the differences stem from how spike sorting handles spike-to-neuron assignments or discards certain spikes altogether.
Such decisions introduce variability and potential bias, as some spikes may be misclassified or incorrectly excluded when they should be associated with neurons. These misassignments can affect downstream decoding performance. The marked point process framework, which models the probability of being at a particular position given a spike waveform, provides a principled alternative that avoids incorrect assignments, as discussed in \cite{Ventura2009}.

Although the percentages in the top row of panel (B) indicate how often the clusterless-full model outperforms the other models, they do not convey the magnitude of the improvement. To address this, we computed the difference in reduction of entropy, absolute error, and root mean square error (RMSE) at each spike time. The second and third rows of Panel (B) summarize these results across datasets. The second row shows the mean difference for each metric in three pairwise comparisons: (1) clusterless full vs. clusterless subset, (2) clusterless subset vs. spike-sorted, and (3) clusterless full vs. spike-sorted. The third row shows the corresponding skewness, showing the distribution of differences across time points. Positive mean differences in all three comparisons suggest that the first model consistently outperforms the second, while positive skewness indicates that large improvements occur more frequently than large declines.

This analysis quantifies both the frequency and the magnitude of improvement in clusterless decoding over spike-sorted decoding. The consistently positive means and right-skewed distributions, most notably in entropy reduction, indicate that the improvement is driven by the inclusion of additional spikes in the full dataset. Since the decoding models are otherwise identical, the gain reflects the extra spiking information rather than differences in model structure.  

These findings show that clusterless decoding consistently recovers information that is lost when spikes are discarded during sorting. In the Discussion, we place these results in the broader context of neural coding and highlight the limitations of conventional spike-sorting pipelines.

\section{Discussion}

In this paper, we developed information-theoretic measures to assess the amount of information carried by spikes in a neural population. These methods quantify the information content of individual spikes both in isolation and in relation to the ongoing history of population activity, that is, their incremental contribution given the information already conveyed by spikes in preceding time steps. While traditional approaches often treat spikes independently, our framework captures how each spike contributes both on its own and in relation to the broader dynamics of the population. When information is measured from spikes in isolation using a flat prior, most spikes appear highly informative, particularly those with high amplitude, which consistently show greater information content than low-amplitude spikes. However, when incremental information is evaluated, the unique contribution of each spike is reflected in the activity of other neurons, the total information content decreases substantially. This reduction reflects the redundancy present in the population. Moreover, the distinction between high- and low-amplitude spikes becomes markedly less pronounced: spikes that appear uninformative in isolation often carry comparable information when considered within the context of the full population activity. These findings help explain why population-level decoders such as clusterless frequently outperform spike sorted approaches.

We applied our information-theoretic framework to address a central question: does the improvement in clusterless decoding primarily stem from a reduction in bias, or from the inclusion of low-amplitude spikes that could not be clustered and were previously assumed to carry little information? Our results suggest that the answer is not straightforward, neither factor consistently dominates. Even within the same animal and recording session, we observed subsets of the data in which performance gains were primarily attributable to bias reduction and others in which the inclusion of low-amplitude spikes was the key contributor. Crucially, these distinctions were only apparent through the use of information-theoretic analysis, underscoring the value of these tools to interpret the sources of decoding performance.

A limitation of this study is that it focuses exclusively on the hippocampus in a single species and relies on a limited set of information-theoretic measures. The observed effects of low-amplitude spikes may reflect properties specific to this system, and a broader generalization will require validation across other brain areas, behavioral contexts, and animal models. In addition, we focused on reduction in entropy and mean squared error to assess information content and decoding performance. While these metrics are commonly used, they do not capture all aspects of neural coding. Finally, our analysis is based on decoding models; therefore, any limitations in these models, such as potential biases from model misclassifications or assumptions, such as how spike features relate to behavior, can bias information estimates. More broadly, it is unlikely that any single metric can fully characterize the contribution of a spike to neural representation. This highlights the need for a more comprehensive analytical framework, one that incorporates multiple complementary measures to probe different aspects of population coding. Interpreting neural information therefore requires a principled alignment between the choice of metric and the specific scientific question under investigation.

Future work might extend this framework to include a broader range of spike features, additional brain systems including sensory and motor areas such as visual and motor cortices, diverse experimental tasks, and a wider set of information-theoretic metrics. For example, Kullback--Leibler (KL) divergence \cite{kullback1997}, which compares full probability distributions, may offer more sensitive evaluations of changes in the encoding structure. Other approaches that directly assess covariability between spiking activity and encoded signals, such as theoretical covariance measures that do not rely on decoding models, could provide complementary insights. These methods may reveal new patterns or reproduce key trends identified in our current analysis. Since no single metric fully captures the contribution of each spike to the neural code, applying multiple complementary measures will be essential to investigate different aspects of the information distribution. One core result of this study, the reduced contrast between high- and low-amplitude spikes under incremental information models, will likely generalize across a range of alternative information-theoretic and statistical metrics. Although our current framework is based on decoding models, similar insights may arise from approaches that do not rely on decoding, such as analyses of spike redundancy or temporal covariance that more directly reflect population structure. Ultimately, the choice of information metric should be guided by the specific scientific question and the nature of the information being investigated. Developing tools that support such targeted and interpretable analyses across diverse neural systems will be essential to advance our understanding of population-level coding.

As neural datasets grow larger and more complex, especially in population level coding, it is important to identify where meaningful information actually resides within neural activity, rather than relying on prior assumptions. Our results challenge the common view that low-amplitude or unsorted spikes are just noise, showing that they can carry distinct and behaviorally relevant information. This highlights the need for analytical tools that quantify the contribution of individual spikes and help reveal structure that might otherwise be discarded. By showing that information content depends not only on spike amplitude or whether a spike can be cleanly sorted, but also on its temporal context and the surrounding population activity, we highlight the need for a broader perspective on neural coding, one that takes into account the full range of spiking activity.

\newpage

\bibliography{references}

@article{Brown1998,
  author    = {Brown, Emery N. and Frank, Loren M. and Tang, Dawn and Quirk, Michael C. and Wilson, Matthew A.},
  title     = {A statistical paradigm for neural spike train decoding applied to position prediction from ensemble firing patterns of rat hippocampal place cells},
  journal   = {The Journal of Neuroscience},
  volume    = {18},
  number    = {18},
  pages     = {7411--7425},
  year      = {1998},
  doi       = {10.1523/JNEUROSCI.18-18-07411.1998}
}

@article{Chen2014,
  author    = {Chen, Zhe and Gomperts, Stephen N. and Yamamoto, Jun and Wilson, Matthew A.},
  title     = {Neural representation of spatial topology in the rodent hippocampus},
  journal   = {Neural Computation},
  volume    = {26},
  number    = {1},
  pages     = {1--39},
  year      = {2014},
  doi       = {10.1162/NECO_a_00538}
}

@inproceedings{Zoltowski2020ML,
  author    = {Zoltowski, David M. and Pillow, Jonathan W. and Linderman, Scott W.},
  title     = {A general recurrent state space framework for modeling neural dynamics during decision-making},
  booktitle = {Proceedings of the 37th International Conference on Machine Learning (ICML)},
  volume    = {119},
  pages     = {11680--11691},
  year      = {2020},
  editor    = {Daumé, Hal III and Singh, Aarti},
  series    = {Proceedings of Machine Learning Research},
  publisher = {PMLR},
  doi       = {10.5555/3524938.3526021}
}

@article{Denovellis2021,
  author    = {Denovellis, Eric L. and Gillespie, Anna K. and Coulter, Michael E. and Sosa, Marielena and Chung, Jason E. and Eden, Uri T. and Frank, Loren M.},
  title     = {Hippocampal replay of experience at real-world speeds},
  journal   = {eLife},
  volume    = {10},
  year      = {2021},
  pages     = {e64505},
  doi       = {10.7554/eLife.64505}
}

@article{Gillespie2021,
  author    = {Gillespie, Anna K. and Astudillo, Oscar and LaFratta, Melissa D. and others},
  title     = {Hippocampal replay reflects specific past experiences rather than a plan for subsequent choice},
  journal   = {Neuron},
  volume    = {109},
  number    = {19},
  pages     = {3149--3163.e6},
  year      = {2021},
  publisher = {Elsevier},
  doi       = {10.1016/j.neuron.2021.07.029}
}

@article{Kloosterman2013,
  author    = {Kloosterman, Fabian and Layton, Stuart P. and Chen, Zhe and Wilson, Matthew A.},
  title     = {Bayesian decoding using unsorted spikes in the rat hippocampus},
  journal   = {Journal of Neurophysiology},
  volume    = {111},
  number    = {1},
  pages     = {217--227},
  year      = {2013},
  doi       = {10.1152/jn.01046.2012},
  pmid      = {24089403},
  pmcid     = {PMC3921373}
}

@article{Quiroga2004,
  author    = {Quiroga, R. Quian and Nadasdy, Z. and Ben Shaul, Y.},
  title     = {Unsupervised spike detection and sorting with wavelets and superparamagnetic clustering},
  journal   = {Neural Computation},
  volume    = {16},
  number    = {8},
  pages     = {1661--1687},
  year      = {2004},
  doi       = {10.1162/089976604774201631}
}

@article{VargasIrwin2007,
  author    = {Vargas-Irwin, Carlos and Donoghue, John P.},
  title     = {Automated spike sorting using density grid contour clustering and subtractive waveform decomposition},
  journal   = {Journal of Neuroscience Methods},
  volume    = {164},
  number    = {1},
  pages     = {1--18},
  year      = {2007},
  doi       = {10.1016/j.jneumeth.2007.03.025},
  pmid      = {17512603},
  pmcid     = {PMC2104515}
}

@article{Fee1996,
  author    = {Fee, Michale S. and Mitra, Partha P. and Kleinfeld, David},
  title     = {Automatic sorting of multiple unit neuronal signals in the presence of anisotropic and non-Gaussian variability},
  journal   = {Journal of Neuroscience Methods},
  volume    = {69},
  number    = {2},
  pages     = {175--188},
  year      = {1996},
  issn      = {0165-0270},
  doi       = {10.1016/S0165-0270(96)00050-7}
}

@article{Lewicki1998,
  author    = {Lewicki, Michael S.},
  title     = {A review of methods for spike sorting: the detection and classification of neural action potentials},
  journal   = {Network: Computation in Neural Systems},
  volume    = {9},
  number    = {4},
  pages     = {R53--R78},
  year      = {1998},
  doi       = {10.1088/0954-898X_9_4_001}
}

@article{Wood2004,
  author    = {Wood, Frank and Black, Michael J. and Vargas-Irwin, Carlos and Fellows, Matthew and Donoghue, John P.},
  title     = {On the variability of manual spike sorting},
  journal   = {IEEE Transactions on Biomedical Engineering},
  volume    = {51},
  number    = {6},
  pages     = {912--918},
  year      = {2004},
  doi       = {10.1109/TBME.2004.826677}
}

@article{Einevoll2012,
  author    = {Einevoll, Gaute T. and Franke, Felix and Hagen, Espen and Pouzat, Christophe and Harris, Kenneth D.},
  title     = {Towards reliable spike-train recordings from thousands of neurons with multielectrodes},
  journal   = {Current Opinion in Neurobiology},
  volume    = {22},
  number    = {1},
  pages     = {11--17},
  year      = {2012},
  doi       ={10.1016/j.conb.2011.10.001},
  issn      = {0959-4388}
}

@article{Ventura2009,
  author    = {Ventura, Val{\'e}rie},
  title     = {Traditional waveform-based spike sorting yields biased rate code estimates},
  journal   = {Proceedings of the National Academy of Sciences},
  volume    = {106},
  number    = {17},
  pages     = {6921--6926},
  year      = {2009},
  doi       = {10.1073/pnas.0901771106}
}

@article{Paulk2022,
  author    = {Paulk, Angelique C. and others},
  title     = {Large-scale neural recordings with single neuron resolution using Neuropixels probes in human cortex},
  journal   = {Nature Neuroscience},
  volume    = {25},
  number    = {2},
  pages     = {252--263},
  year      = {2022},
  doi       = {10.1038/s41593-021-00997-0}
}

@article{Lefebvre2016,
  author    = {Lefebvre, Baptiste and Yger, Pierre and Marre, Olivier},
  title     = {Recent progress in multi-electrode spike sorting methods},
  journal   = {Journal of Physiology-Paris},
  volume    = {110},
  number    = {4},
  pages     = {327--335},
  year      = {2016},
  doi       = {10.1016/j.jphysparis.2017.02.005},
  pmid      = {28263793},
  pmcid     = {PMC5581741}
}

@article{Todorova2014,
  author  = {Todorova, Stiliyan and Sadtler, Patrick and Batista, Aaron and Chase, Steven and Ventura, Valerio},
  title   = {To Sort or Not to Sort: The Impact of Spike-Sorting on Neural Decoding Performance},
  journal = {Journal of Neural Engineering},
  volume  = {11},
  number  = {5},
  pages   = {056005},
  year    = {2014},
  doi     = {10.1088/1741-2560/11/5/056005},
  pmid    = {25082508},
  pmcid   = {PMC4454741}
}

@article{Deng2015,
  author  = {Deng, Xinyi and Liu, Daniel F. and Kay, Kenneth and Frank, Loren M. and Eden, Uri T.},
  title   = {Clusterless Decoding of Position from Multiunit Activity Using a Marked Point Process Filter},
  journal = {Neural Computation},
  volume  = {27},
  number  = {7},
  pages   = {1438--1460},  
  year    = {2015},
  doi     = {10.1162/NECO_a_00744},
  issn    = {1530-888X}
}

@article{Deng2016,
  author  = {Deng, Xinyi and Liu, Daniel F. and Karlsson, Mattias P. and Frank, Loren M. and Eden, Uri T.},
  title   = {Rapid Classification of Hippocampal Replay Content for Real-Time Applications},
  journal = {Journal of Neurophysiology},
  volume  = {116},
  number  = {5},
  pages   = {2221--2235},
  year    = {2016},
  publisher = {American Physiological Society},
  doi     = {10.1152/jn.00151.2016}
}

@article{Tao2018,
  author = {Tao, Long and Weber, Karoline E. and Arai, Kensuke and Eden, Uri T.},
  title = {A Common Goodness-of-Fit Framework for Neural Population Models Using Marked Point Process Time-Rescaling},
  journal = {bioRxiv},
  year = {2018},
  note = {bioRxiv preprint. Available at \url{https://doi.org/10.1101/265850}}
}

@article{Kay2020,
  author    = {Kay, K. and Chung, J. E. and Sosa, M. and Schor, J. S. and Karlsson, M. P. and Larkin, M. C. and Liu, D. F. and Frank, L. M.},
  title     = {Constant Sub-second Cycling between Representations of Possible Futures in the Hippocampus},
  journal   = {Cell},
  volume    = {180},
  number    = {3},
  pages     = {552--567.e25},
  year      = {2020},
  doi       = {10.1016/j.cell.2020.01.014}
}

@article{shannon1948,
  author  = {Shannon, Claude E.},
  title   = {A Mathematical Theory of Communication},
  journal = {Bell System Technical Journal},
  volume  = {27},
  number  = {3},
  pages   = {379--423},
  year    = {1948},
  doi     = {10.1002/j.1538-7305.1948.tb01338.x}
}

@article{Borst1999,
  author    = {Borst, Alexander and Theunissen, Frédéric E.},
  title     = {Information theory and neural coding},
  journal   = {Nature Neuroscience},
  volume    = {2},
  number    = {11},
  pages     = {947--957},  
  year      = {1999},
  doi       = {10.1038/14731},
  issn      = {1097-6256}
}

@article{Hurwitz1975,
  author    = {Hurwitz, Henry},
  title     = {Entropy Reduction in Bayesian Analysis of Measurements},
  journal   = {Physical Review A},
  volume    = {12},
  pages     = {698--706},
  year      = {1975},
 publisher = {American Physical Society},
  doi       = {10.1103/PhysRevA.12.698}
}

@book{Rieke1997,
  author    = {Rieke, Fred and Warland, David and de Ruyter van Steveninck, Rob and Bialek, William},
  title     = {Spikes: Exploring the Neural Code},
  year      = {1997},
  publisher = {MIT Press},
  address   = {Cambridge, MA}
}

@article{Eden2004,
  author    = {Eden, Uri T. and Frank, Loren M. and Barbieri, Riccardo and Solo, Victor and Brown, Emery N.},
  title     = {Dynamic analysis of neural encoding by point process adaptive filtering},
  journal   = {Neural Computation},
  volume    = {16},
  number    = {5},
  pages     = {971--998},
  year      = {2004},
  doi       = {10.1162/089976604773135069}
}

@article{brenner2000synergy,
    author  = {Brenner, Naama and Strong, Steven P. and Koberle, Roland and Bialek, William and de Ruyter van Steveninck, Rob R.},
    title   = {Synergy in a Neural Code},
    journal = {Neural Computation},
    year    = {2000},
    volume  = {12},
    number  = {7},
    pages   = {1531--1552},
    doi     = {10.1162/089976600300015259},
    pmid    = {10935917}
}

@article{Eden2008,
  author  = {Eden, Uri T. and Brown, Emery N.},
  title   = {Continuous-Time Filters for State Estimation from Point Process Models of Neural Data},
  journal = {Statistica Sinica},
  volume  = {18},
  number  = {4},
  pages   = {1293--1310},
  year    = {2008},
  pmid    = {22065511},
  pmcid   = {PMC3208353}
}

@techreport{Karlsson2015,
  author      = {Karlsson, Mattias and Carr, Margaret and Frank, Loren M.},
  title       = {Simultaneous extracellular recordings from hippocampal areas CA1 and CA3 (or MEC and CA1) from rats performing an alternation task in two W-shaped tracks that are geometrically identical but visually distinct},
  year        = {2015},
  institution = {CRCNS.org},
  type        = {Dataset},
  url         = {http://dx.doi.org/10.6080/K0NK3BZJ},
  doi         = {10.6080/K0NK3BZJ}
}

@article{frank2004,
  author    = {Frank, Loren M. and Stanley, Garrett B. and Brown, Emery N.},
  title     = {Hippocampal Plasticity Across Multiple Days of Exposure to Novel Environments},
  journal   = {Journal of Neuroscience},
  volume    = {24},
  number    = {35},
  pages     = {7681--7689},
  year      = {2004},
  doi       = {10.1523/JNEUROSCI.1958-04.2004},
  pmid      = {15342735},
  pmcid     = {PMC6729632}
}

@article{karlsson2008,
  author    = {Karlsson, Mattias P. and Frank, Loren M.},
  title     = {Network Dynamics Underlying the Formation of Sparse, Informative Representations in the {Hippocampus}},
  journal   = {Journal of Neuroscience},
  volume    = {28},
  number    = {52},
  pages     = {14271--14281},
  year      = {2008},
  doi       = {10.1523/JNEUROSCI.4261-08.2008},
  pmid      = {19109508},
  pmcid     = {PMC2632980}
}

@article{wilson1994reactivation,
  author    = {Wilson, M. A. and McNaughton, B. L.},
  title     = {Reactivation of hippocampal ensemble memories during sleep},
  journal   = {Science},
  volume    = {265},
  number    = {5172},
  pages     = {676--679},
  year      = {1994},
  doi       = {10.1126/science.8036517},
  pmid      = {8036517}
}

@article{karlsson2009awake,
  author    = {Karlsson, M. P. and Frank, L. M.},
  title     = {Awake replay of remote experiences in the hippocampus},
  journal   = {Nature Neuroscience},
  volume    = {12},
  number    = {7},
  pages     = {913--918},
  year      = {2009},
  doi       = {10.1038/nn.2344},
  pmid      = {19525943},
  pmcid     = {PMC2750914}
}

@article{davidson2009,
  author    = {Davidson, T. J. and Kloosterman, F. and Wilson, M. A.},
  title     = {Hippocampal replay of extended experience},
  journal   = {Neuron},
  volume    = {63},
  number    = {4},
  pages     = {497--507},
  year      = {2009},
  doi       = {10.1016/j.neuron.2009.07.027},
  pmid      = {19709631},
  pmcid     = {PMC4364032}
}

@book{kullback1997,
  author    = {Solomon Kullback},
  title     = {Information Theory and Statistics},
  publisher = {Courier Corporation},
  year      = {1997},
  address   = {Paris, France}
}

\end{document}